\documentclass[journal]{IEEEtran}
\usepackage{color}
\usepackage{graphicx}
\usepackage{tabularx}
\usepackage{booktabs}
\usepackage{multirow}
\usepackage{amsmath}
\usepackage{array}
\usepackage{subfigure}
\usepackage{xcolor,cite,etoolbox}
\makeatletter
\pretocmd\@bibitem{\color{black}\csname keycolor#1\endcsname}{}{\fail}
\newcommand\citecolor[1]{\@namedef{keycolor#1}{\color{red}}}
\makeatother
\begin{document}

% \title{Molecular Communication Meets Virus: Modeling, Detection, and Prediction}
%\title{\LARGE{Wireless Molecular Communications Empowered Viral Infections Research: Modeling, Detection, and Prediction}}

\title{\LARGE{Empowering Nanoscale Connectivity through Molecular Communication:\\ A Case Study of Virus Infection}}

\author{Xuan Chen, Yu Huang, Miaowen Wen,~\emph{Senior Member,~IEEE,}  Shahid Mumtaz,~\emph{Senior Member,~IEEE,}  \\
Fatih Gulec, Anwer Al-Dulaimi,~\emph{Senior Member,~IEEE,} and Andrew W. Eckford,~\emph{Senior Member,~IEEE} \\

\thanks{
This work was supported in part by the National Natural Science Foundation of China under Grants 62471183, 62301173, and 62201161, and in part by the NSERC Discovery grant RGPIN-2016-05288. \emph{(Corresponding author: Miaowen Wen.)}}

\thanks {X. Chen and Y. Huang are with the Research Center of Intelligent Communication Engineering, School of Electronics and Communication Engineering, Guangzhou University, Guangzhou 510006, China (Email: \{eechenxuan, yuhuang\}@gzhu.edu.cn).

     M. Wen is with the School of Electronic and Information Engineering, South China University of Technology, Guangzhou 510640, China (Email: eemwwen@scut.edu.cn).

S. Mumtaz is with Nottingham Trent University, Nottingham, UK (Email: Dr.Shahid.mumtaz@ieee.org).

F. Gulec is with the School of Computer Science and Electronic Engineering, University of Essex, UK (Email: f.gulec@essex.ac.uk).

A. Al-Dulaimi is with Veltris, Toronto L6W 3W8, Canada, and also
with Zayed University, Abu Dhabi, UAE (Email: anwer.aldulaimi@ieee.org).

A. W. Eckford is with the Department of Electrical Engineering and Computer Science, York University, Toronto M3J 1P3, Canada (Email: aeckford@yorku.ca).

}}

%\markboth{Submitted to IEEE Communications Magazine}%

\maketitle

\begin{abstract}
The Internet of Bio-Nano Things (IoBNT), envisioned as a revolutionary healthcare paradigm, shows promise for epidemic control.
This paper explores the potential of using molecular communication (MC) to address the challenges in constructing IoBNT for epidemic prevention, specifically focusing on modeling viral transmission, detecting the virus/infected individuals, and identifying virus mutations.
First, the MC channels in macroscale and microscale scenarios are discussed to match viral transmission in both scales separately.
Besides, the detection methods for these two scales are also studied, along with the localization mechanism designed for the virus/infected individuals.
Moreover, an identification strategy is proposed to determine potential virus mutations, which is validated through simulation using the ORF3a protein as a benchmark.
Finally, open research issues are discussed.
In summary, this paper aims to analyze viral transmission through MC and combat viral spread using signal processing techniques within MC.

\end{abstract}


\IEEEpeerreviewmaketitle

\vspace{-0.8cm}
\section{Introduction}

The Internet of Things (IoT) enables seamless connectivity from macro to nanoscale. One notable extension of IoT is the Internet of Bio-Nano Things (IoBNT), which incorporates nanotechnology into the biological domain \cite{akyildiz2015internet}. With nanoscale connectivity and high biocompatibility, IoBNT holds promise in combating infectious diseases by facilitating early detection, real-time monitoring, and targeted treatment \cite{akyildiz2020panacea}. This exploration of IoBNT has gained significant attention, especially in the post-COVID-19 era. Yet, to effectively use IoBNT in epidemic prevention, two critical challenges need to be addressed: i) establishing an efficient communication paradigm between devices, and ii) gaining a fresh understanding of infectious diseases. The first challenge arises from the fact that certain common scenarios in IoBNT, such as within the human body or in nanonetworks, are wave-unfriendly, making traditional communication difficult to apply.
The second challenge involves knowing the transmission patterns of viral infection to design a detection and warning system within the IoBNT.
As a nature-inspired communication paradigm, molecular communication (MC) presents a promising solution to address these challenges. On one hand, by transmitting information through molecules instead of waves, MC has the potential to support robust communication in wave-unfriendly environments. Besides, recent research suggests the feasibility of using MC to capture the key factors of infectious diseases and reveal their spread mechanism \cite{Visu_model_MT_TMBMC}.
Several great works have been proposed to explore using MC for enabling communication in the IoBNT, while research on the second challenge faced by IoBNT is still limited.
Hence, this paper takes virus infection as an example to show how MC facilitates IoBNT.

Driven by the COVID-19 pandemic, there has been a surge of research interest in viral infection research empowered by MC, or saying, understanding viral infection from the view of Information and Communication Technology (ICT).
Among these studies, most researchers focused on modeling~pathogen transmission, including particle dispersion in the air and virus spread within the body \cite{Khalid.AerosolMCmodel.20}. In general, MC can be divided into macroscale and microscale, corresponding to viral transmission scenarios for interpersonal and intra-body infections \cite{Viral_propagation_in_respiratory}. Besides, the detection scheme to monitor the suspected cases has been developed for the macroscale environment based on the proposed virus propagation models, illuminating an ICT solution for epidemic control. Unfortunately, when modeling viral transmission on the macroscale, it is generally assumed that two static humans are considered while ignoring the interaction with others and their mobility. Similar issues arise when simulating viral transmission on the microscale. Moreover, the detection range is limited to macroscopic scenarios, which also restricts the development space of MC in virus research. Finally, the diversity of viruses arising from evolution and natural selection over time poses a great challenge to virus research.

\begin{figure*}[htbp]
\begin{center}
  \includegraphics[width=4.85in]{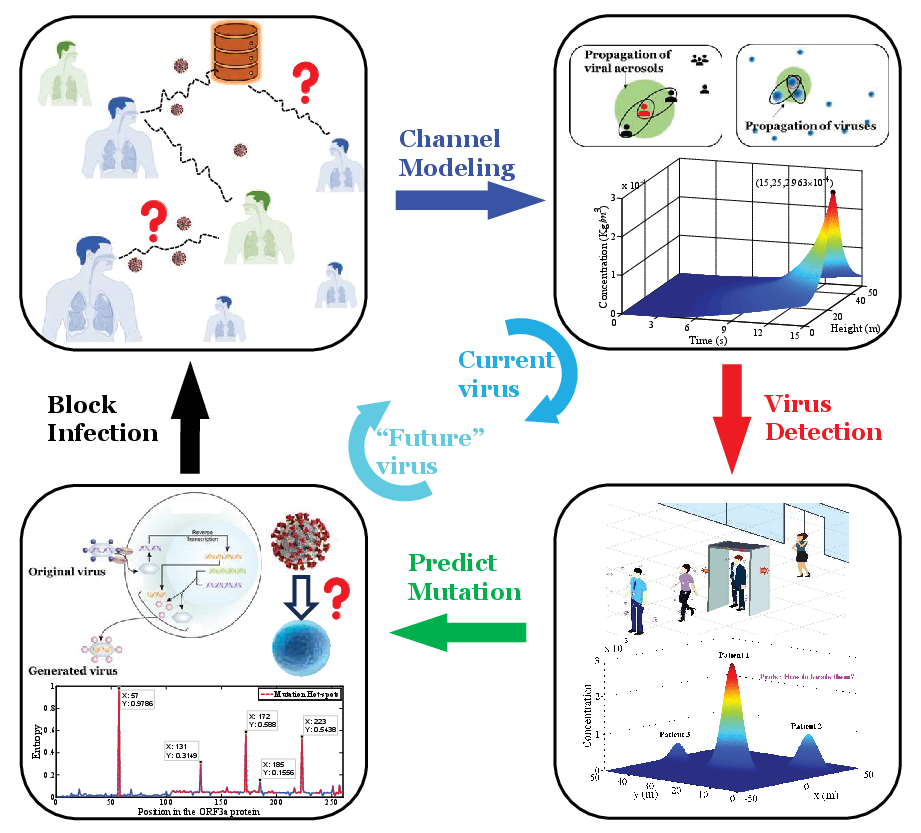}
  \end{center}
  \vspace{-0.2cm}
  \caption{Schematic diagram for the viral infection research.}
    \vspace{-0.2cm}
  \label{figure-System_model_total}
\end{figure*}

Against this background, the following questions arise: how to accurately model viral transmission; how to effectively detect the virus/infected individuals; and how to reliably identify the mutation of viruses. The inner connection between these key questions is depicted in Fig.~\ref{figure-System_model_total}. In this paper, we are committed to giving a feasible insight into these issues.

\vspace{-0.1cm}
\section{Modeling of Viral Transmission via Molecular Communication}

Transmission models play a significant role in the study of viral infections. In the literature, the spread of a virus can be viewed as signaling, i.e., transmitter (emitting a virus), channel (spreading a virus), and receiver (invading a host). Owing to this similarity, we focus on how to model the viral infection with the aid of MC in this section.

\subsection{Viral Transmission Model}

In the case of individuals, viral transmission can be viewed as an interaction between an infected and a susceptible; while in the human body, it can be viewed as an interaction between a virus and a healthy cell. Per the multi-scale nature of MC, virus propagation can be modeled at both macro- and micro-scales, as shown in Fig.~\ref{figure-Viral Propagation in vitro}.

\subsubsection{Viral Propagation at the Macroscale}

\begin{figure*}[htbp]
\begin{center}
  \includegraphics[width=5.25in]{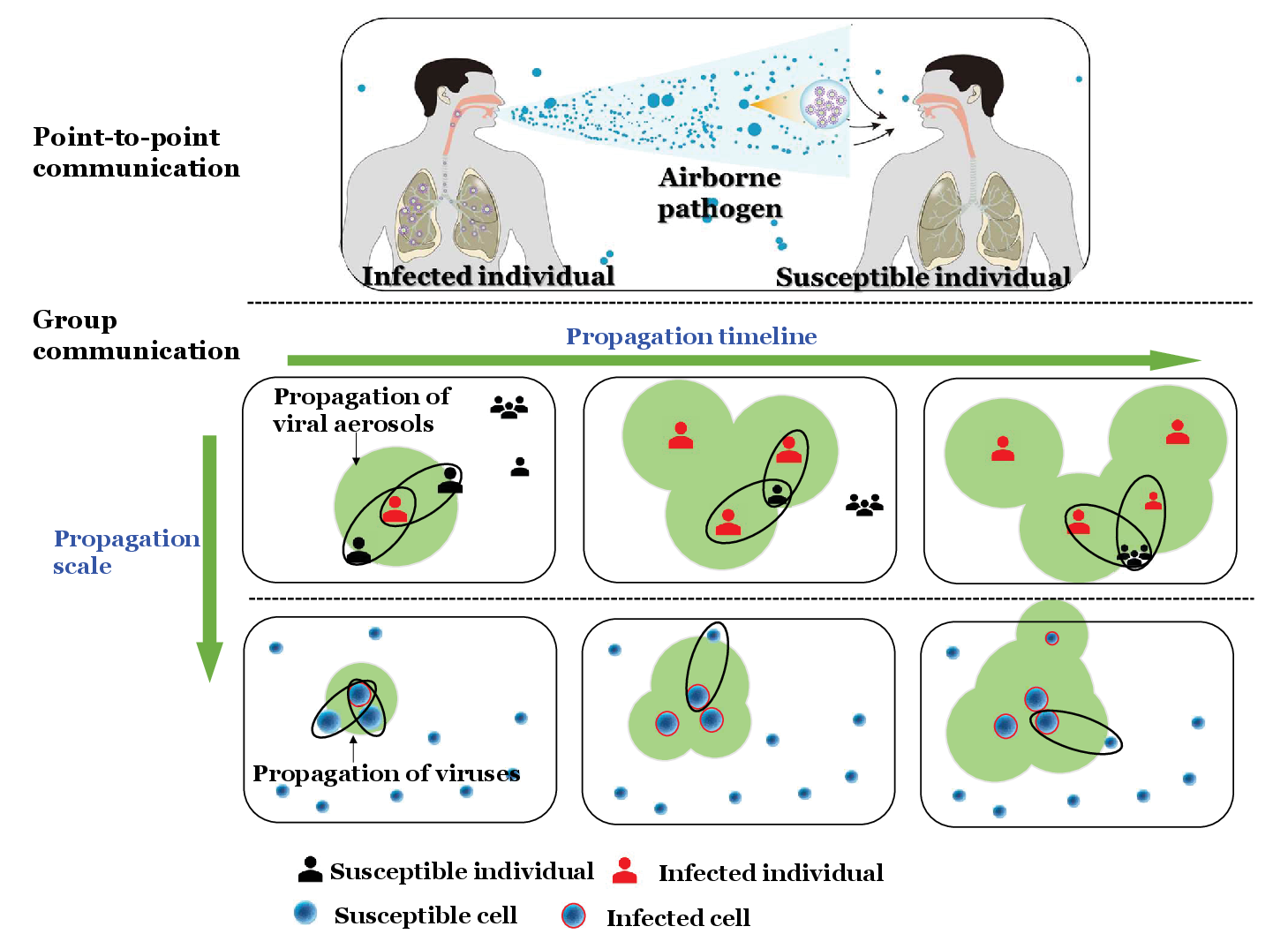}
  \end{center}
      \vspace{-0.3cm}
  \caption{A viral transmission model. Top subfigure: Point-to-point transmission between two static individuals. Bottom subfigure: An epidemic transmission scenario from the macroscale to the microscale.}
        \vspace{-0.2cm}
  \label{figure-Viral Propagation in vitro}
\end{figure*}

Airborne transmission is the main avenue for human-to-human virus spread, and thereby we take it as an example to explore viral propagation at the macroscale. Varied individual states and environments influence transmission modeling. For clarity, we mainly discuss three categories of transmission models: i) between two static individuals; ii) between a moving and static individual; and iii) within a population.

As shown in Fig.~\ref{figure-Viral Propagation in vitro}, a viral propagation model between two individuals can be regarded as a point-to-point transmission. Herein, the transmitter of viruses is the infected, the information conveyed is the ``infection" action, and the receiver is the susceptible. Aerosol propagation in the air is essentially a diffusion process, and its concentration follows Fick's second law. For clarity, we define ${c({\bf{r}},t)}$~as~the~aerosol concentration at a position ${\bf{r}} = [x,y,z]$ and time~$t$.
Per Fick's second law, the rate of change in concentration over time $\frac{{\partial c({\bf{r}},t)}}{{\partial t}}$ is proportional to the spatial second derivatives of the concentration, with the diffusivity $D$ as the coefficient. When airflow is also involved in aerosol propagation, the diffusion problem is transformed into an advection-diffusion equation. This equation includes the diffusion term $ \frac{{{\partial ^2}c({\bf{r}},t)}}{{\partial {{\bf{r}}^2}}}$ and the advection term $ v\frac{{\partial c(\mathbf{r},t)}}{{\partial \mathbf{r}}}$ to calculate the concentration $c(\mathbf{r},t)$, where $v$ is the airflow velocity. Yet, when deriving the virus-bearing airborne concentration distribution, we should note~that
\begin{itemize}
    % \item The infected can be viewed as an instantaneous source with an impulsive emission or a continuous source with a square-wave emission, according to different human behaviors, such as coughing and breathing. The airborne-releasing rate is proportional to the severity of symptoms in the infected, ranging from mild to severe.
    \item The infected can act as either an instant or continuous source according to different human behaviors, and the airborne-releasing rate is proportional to the severity of symptoms in the infected, ranging from mild to severe.
    \item A larger diffusivity, typically from $10^2$ to $10^{10}$ times larger, could be approximately used to simulate the impact of air turbulence, due to thermal effects, people's movements, wind, etc., since
    $c(\mathbf{r},t)$ cannot be solved well when the diffusivity $D$ varies with respect to $({\bf{r}},t)$.
    \item In the presence of external forces such as wind, flow, and gravity, the velocity $v$ is assumed to be revised as $v({\bf{r}},t)$. Similarly, $v({\bf{r}},t)$ will make $c({\bf{r}},t)$ harder to solve, and an~average velocity from multiple experiments is preferred.
\end{itemize}
The above analysis illustrates how to obtain the concentration space-time distribution for a scenario with two static individuals. While considering a moving infected person, the typical mobility models, such as the random direction model, random walk model, and random waypoint model, can first be used to describe the movement of mobile individuals. For each snapshot, a moving infected person can be viewed as a static one. Here, we can find that the aerosol concentration at the current time is affected by all past values of source position and release rate, and thereby its mathematical form should be a time-accumulation function about $c({\bf{r}},t)$~\cite{zhao2006detecting}.

We further consider an epidemic dissemination system, as shown in Fig.~\ref{figure-Viral Propagation in vitro}. Viral transmission between multi-individuals can be defined as an MC mobile network, where humans function as mobile users in two states: infected and susceptible \cite{gulec2022mobile}.
In this paper, the susceptible-infectious (SI) model is used to analyze viral propagation.
% This is because (i) it is nearly impossible for humans to realize a recovery from the infected state \textit{in a short period}; (ii) simpler models are more likely to obtain a closed-form channel model, providing the basis for detection or localization of the infected.
Compared to the two-person case, the scenario considered is an extension from a point source to a distributed source.
% In other words, an MC mobile network can be easily equivalent to multiple parallel point-to-point transmission systems for analysis.
Therefore, the aerosol concentration should be formulated as a location-accumulation function about $c({\bf{r}},t)$ for a given~$t$. $c({\bf{r}},t)$ is highly dependent on human activity for case ii) and case iii), and thereby sampling enough times is a must to derive the physical model for the spatial and temporal concentration distribution of the diffused aerosol from the moving source. Similarly, it is reasonably assumed that a coherence time holds for case iii), since it takes some time for a newly infected person to be contagious. During this time, the derived $c({\bf{r}},t)$ is only related to the location and symptoms of the infected and not to the number of the infected person. When all factors are averaged, the derived $c({\bf{r}},t)$ has the potential to remain unchanged during the coherence time.

\begin{figure*}[htb]
    \centering
    \subfigure[Velocity = 0 m/s]{
        \includegraphics[width=2.2in]{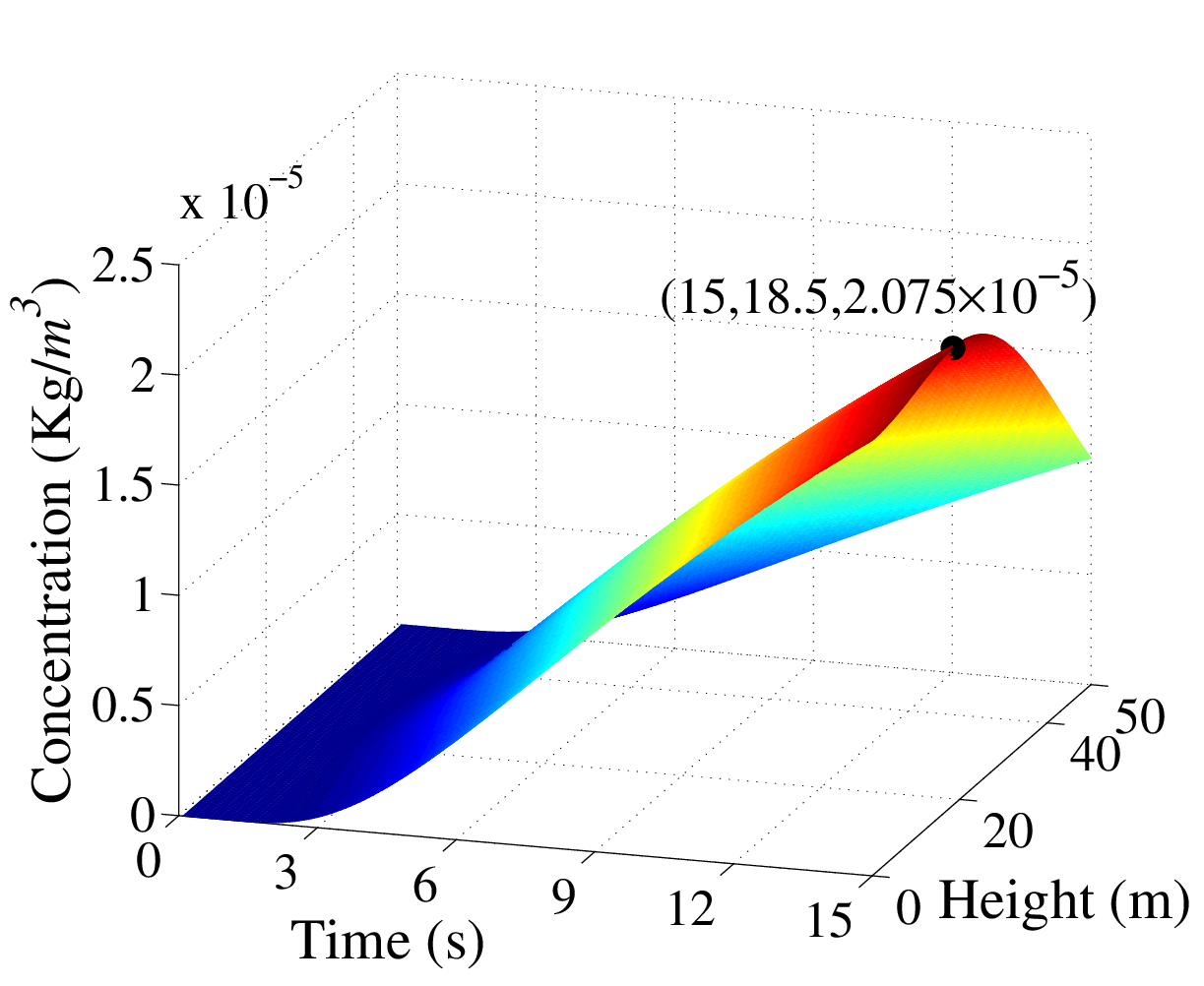}
    \label{CIR_v=0}
    }
        \subfigure[Velocity = 1 m/s]{
	\includegraphics[width=2.2in]{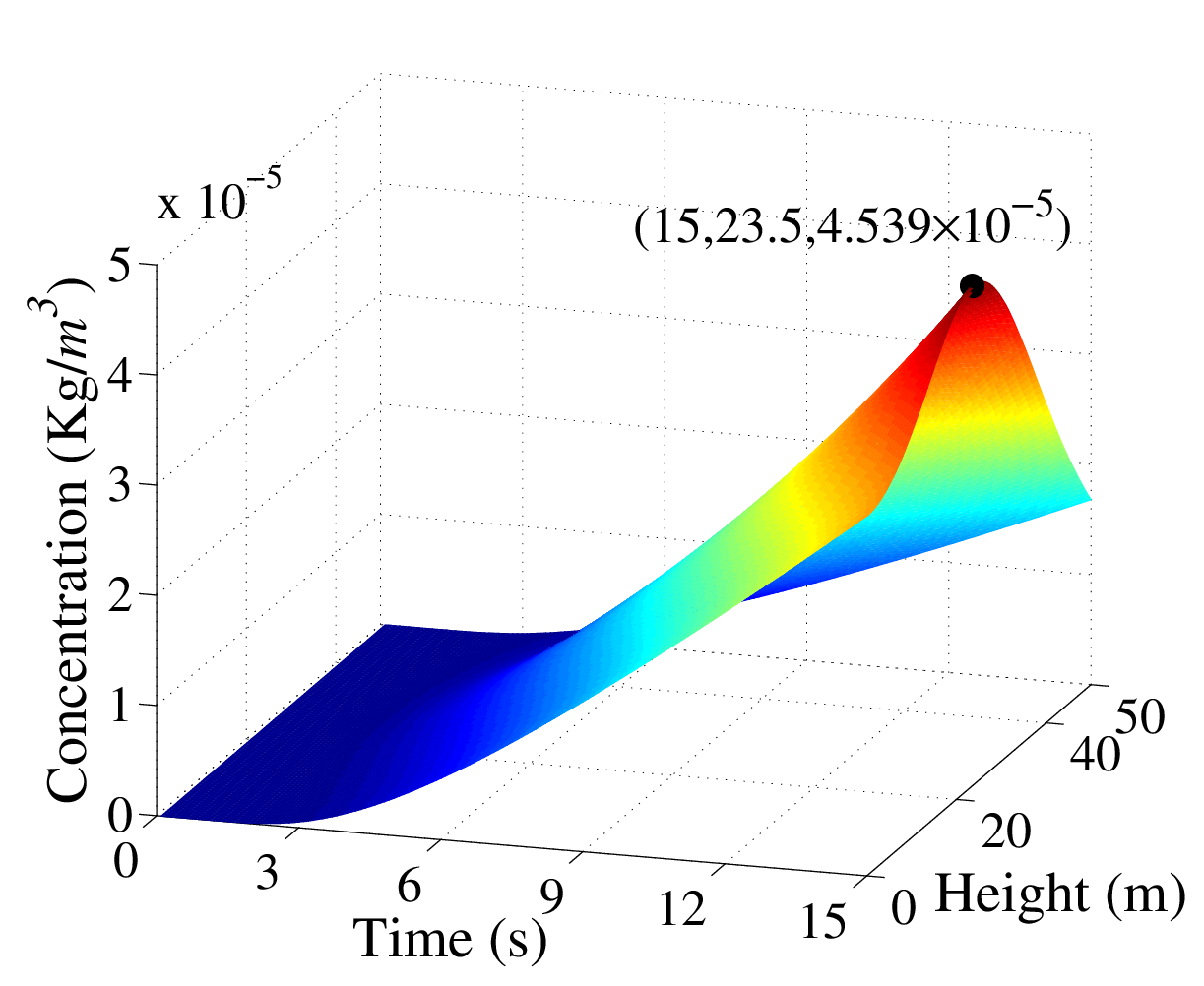}
	\label{CIR_v=1}
	}
	    \subfigure[Velocity = 2 m/s]{
	\includegraphics[width=2.2in]{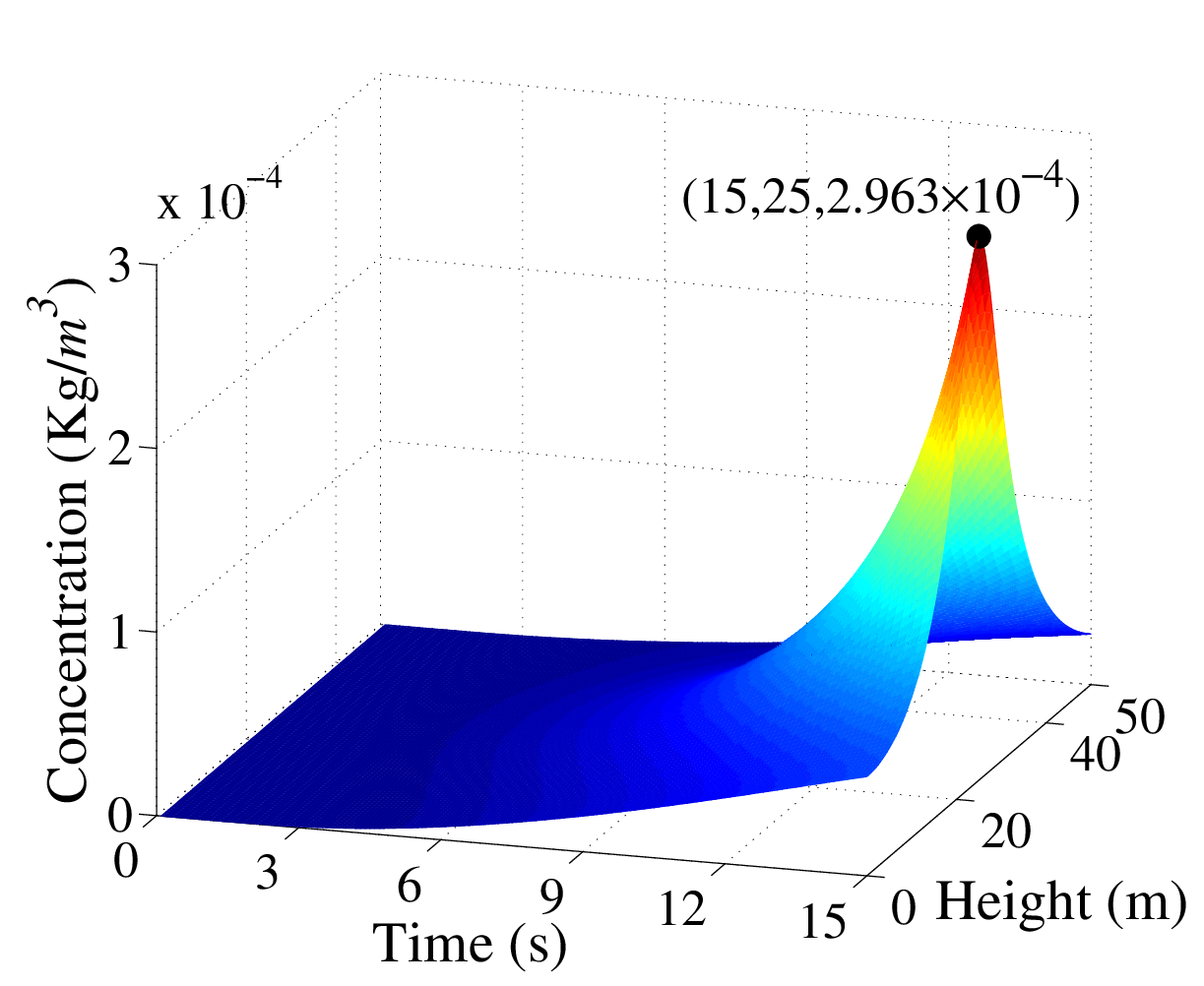}
	\label{CIR_v=3}
	}
 \vspace{-0.15cm}
    \caption{Virus-bearing airborne concentration distribution for case i) and case ii) at the macroscale: a) the infected is static; b) the moving speed for the infected is 1 m/s; c) the moving speed for the infected is 2 m/s. It is assumed that the initial position of the infected is at the point (0, 0, 25) m (to simulate a scene where an infected person stands on a high floor), the diffusivity is 40 $\text{m}^2/\text{s}$, and their breathing is considered a continuous source with a constant releasing rate of 1 Kg/s. Besides, the infected is moving along a straight line that parallels to $x$ axis at a constant height of 25 m, and the susceptible is located at the $x-z$ plane with $x=35$ m and $z \in \{0,50\}$ m.}
     \vspace{-0.3cm}
    \label{CIR_description}
\end{figure*}

In Fig.~\ref{CIR_description}, we illustrate the concentration distribution for the considered viral transmission.
% It is assumed that the initial position of the infected is at the point (0, 0, 25) m (to simulate a scene where an infected person stands on a high floor), the diffusivity is 40 $\text{m}^2/\text{s}$, and their breathing is considered a continuous source with a constant releasing rate of 1 Kg/s. Besides, the infected is moving along a straight line that parallels to $x$ axis at a constant height of 40 m, and the susceptible is located at the $x-z$ plane with $x=30$ m and $z \in \{0,50\}$ m.
% In Fig. 2, we show the virus-bearing airborne concentration with respect to time and height.
As expected, Fig.~\ref{CIR_description} presents clearly that the most susceptible position is around the height of 25 m, paralleling the height of the infected. With an increasing moving speed for the infected, the aerosol will gather at the target height faster, thus leading to a much faster spread of the virus.

\subsubsection{Viral Propagation at the Microscale}

Considering that a virus invades the human body, it is time to analyze the viral propagation at the microscale. Herein, the transmitter and receiver are the infected and susceptible cells, respectively, while the information carrier is a newly generated virus. Note that generally, the respiratory channel and the cutaneous vascular channel are preferred to be discussed in the literature. For a case with two static cells, the calculation for the virus concentration is similar to that derived at the macroscale. While involving the mobile cells, we need to notice that the cell's mobility comes from cell diffusion or some flow in the human body, such as the airflow in the respiratory tract and the blood flow in the cutaneous vascular channel. Therefore, this mobility can be predicted to describe the movement of cells, compared with a macroscale MC mobile network. Besides, its boundary condition from the propagation channel needs to be added when calculating the virus distribution.
% For example, the boundary condition for the considered channels is similar to the pipeline.
Here, the channel models by solving the diffusion equation described previously under different environment assumptions can be obtained.

\vspace{-0.2cm}
\subsection{Duality Between the MC Model and Viral Spread Model}

\begin{table*}[htbp]
    \centering
    \caption{Comparison Between actual MC channel model and viral transmission model}
    \begin{tabular}{ccccccc}
   Type & Transmission scenario  & Transmitter & Information carrier & Carrier flow & Receiver&Boundary\\
        \hline
        \hline
  \multirow{3}{*}{Viral propagation}&Free space  & Mouth/Nose & Droplet & Air flow & Bio-sensor/Human&No\\
   & Respiratory & Nose & Droplet & Mucus flow & Host cells&Pipeline\\
   & Vascular & Animals' bites & Saliva & Blood flow & Host cells&Pipeline  \\
        \hline
        \hline
   \multirow{3}{*}{MC channel}&Free space & Sprayer & Ethanol droplets & No & Alcohol sensor &No  \\
    &Gas pipeline & Gas dispenser & Acetone molecules & Nitrogen flow & Mass spectrometer & Pipeline  \\
    &Liquid pipeline & Sprayer & Magnetic nano-particles & Water flow & Magnetic susceptibility meter&Pipeline
    \end{tabular}
      \label{table-mapping between MC and viral propagation}
\end{table*}

In this subsection, the duality between the actual MC model and the virus transmission model is described. Typical MC models used in experiments and viral propagation models are compared in Table~I. We find that a notable similarity between these models is the channel composition. Besides, the similarity between these two models can be reflected in their performance metrics:
\begin{itemize}
    \item {\bf{\emph{Effectiveness:}}} It can be measured by mutual information. Contrary to data communication systems, mutual information should be minimized as much as possible in a viral propagation model to lower the infection rate \cite{Mutual_information_meet_Shannon}.
    \item {\bf{\emph{Reliability:}}} It can be described by error probability, which should be maximized for viral transmission to reduce the possibility of the virus binding to the correct receptor.
\end{itemize}
Applying MC in viral spread allows us to understand transmission mechanisms through an end-to-end modeling approach, offering a physical model for subsequent applications like detecting or localizing the infected.

% \begin{figure*}[t]
% \begin{center}
%    \includegraphics[width=5.5in]{Fig4.eps}
%    \end{center}
%   \caption{Real-time virus detection: a) macroscale scenario; b) microscale scenario.}
%   \label{Fig5}
% \end{figure*}

\section{Viral Detection via Molecular Communication}

Viral detection is crucial for infection control and therapeutic strategies. Yet, common detection methods involve tedious sample processing, costly detection equipment, and eligible personnel operation~\cite{Gao_invitrobiosensing_2022}.
As mentioned above, MC systems can exhibit real-time molecular signals in response to virus invasions, thus holding promising potential for the rapid detection of infected individuals.

\vspace{-0.25cm}
\subsection{Detection for Infected Individuals}

Based on the channel environment scale, we classify the detection of infected individuals into macroscale and microscale scenarios.

\subsubsection{Macroscale Scenario}

Taking COVID-19 as an example, MC detects the virus based on the particle transmission nature of SARS-CoV-2, where infected individuals release particles carrying viruses through breathing, speaking, and so on. Here, the state of an individual is binary, termed healthy or infected, analogous to on-off keying (OOK) in communication. This enlightens the researchers to collect the particle samples as an alternative indicator for COVID-19 detection, as they are continuously emitted and easily captured at the MC receiver~\cite{Khalid.AerosolMCmodel.20}. Herein, biosensors offer a promising option~for~MC receivers to achieve particle collection and detection, thanks to their high sensitivity, specificity, and potential for real-time detection.

For clarity, an example of viral detection is depicted, where it is assumed that a group of people stands in a queue at the entrance, whose infected status is unknown. Therefore, the MC receiver is used to collect the ambient air sample and then determine the corresponding infection status. This process is a classical OOK demodulation, and it can be done in a symbol-wise or sequence-wise manner, considering the accuracy and complexity requirement. Notice that this detection means depends on an accurate channel model, which can be varied among the infected individuals due to their intrinsic differences, such as height and vital capacity.
Once the channel model is available, coherent detection methods can be used to determine whether infected or not. Non-coherent detection approaches are preferred when the channel modeling process is unavailable. The current detection method of airborne transmission is finished via binary hypothesis testing according to a threshold that represents the strength of the immune system~\cite{Khalid.AerosolMCmodel.20}.

\subsubsection{Microscale Scenario}

Thanks to the progress in design, fabrication, and propulsion, the advancement of nanotechnology holds great promise for artificial nanomachines~\cite{Gao_invitrobiosensing_2022}. Inspired by cellular networks, researchers propose that a swarm of nanomachines may mimic their natural counterparts to form the nanonetwork via the MC protocol for information exchange. In this case, nanomachines can coordinate with each other to accomplish complex tasks, unlocking their potential to achieve disruptive applications such as real-time health status monitoring and diagnosis. Thus, the presence of analytes, such as cancer cell, viruses, and their critical biomarkers, can be monitored and sensed by the nanomachines in vivo. The detection results can be remotely transmitted to external devices, or fed back to the nanomachines for next-step operations, e.g., combating diseases or viral infections via target drug delivery, while minimizing drug overdosing.

% Fig.~4~(b) shows the case that the specific nanomachine patrols in the vessel, distinguishing the cancer cell from the normal red blood cell via the MC mechanism.

\vspace{-0.2cm}
\subsection{Localization for Infected Individuals}

Detection of infected individuals can be achieved by biological detection means like PCR or analysis via MC. However, locating infected individuals from a spacious area, compared to detecting known individuals, remains a challenging task.
Likewise, for macro- and micro-scales, the infected entities to be located are humans and cells, respectively. Herein, the localization of the infected can be transformed into abnormality localization (AL) in the MC field, where the infected entity is the abnormality. MC-based AL is considered an emerging approach to providing a bio-compatible and energy-efficient AL solution.
A comprehensive survey on the MC-based AL schemes has been presented in \cite{MC_based_AL}. According to the literature, we find that a unified propagation model is crucial for optimally designing MC receiver arrays to estimate the position of unknown infected entities. Based on these models, it is promising to develop maximum likelihood or other low-complexity algorithms to estimate the biochemical parameters (location, release intensity, etc.) of infected subjects.

\section{Prediction of Viral Evolution}

% Without mutation, there would be no evolution. Over time, viruses mutate and evolve like living cells, giving rise to new variants. These mutations can potentially bring viruses a survival advantage, such as being more lethal and contagious. Here, it is crucial to early identify possible viral mutations and then block their rapid spread. Currently, predictions of possible variants are done in the lab, which are precise but complex and time-consuming. In this context, the analysis method with the aid of ICT is presented to overcome the above limitations.

Without mutation, there would be no evolution. Over time, mutations lead to new variants, potentially providing survival advantages, such as being more lethal and contagious. Therefore, early identification of viral mutations is crucial to curb rapid spread. Lab predictions are precise but complex and time-consuming. Here, an ICT-aided analysis method is introduced to address these limitations.

\vspace{-0.1cm}
\subsection{Reproductive Model of Viruses}

To characterize the evolutionary process, i.e., transmitting information from generation to generation, we abstract the transmission of genetic material during reproduction as an MC system, as shown in the lower left part of Fig.~\ref{figure-System_model_total}. Specifically, transceivers represent the original virus entering the cell and the newly generated virus released from the infected cell; the information carrier is the gene sequence; and the channel is the replication, translation, and transcription process inside the invaded cell, in which the noise comes from the transmission error and possible mutation. Unlike real communication systems, equivalent ones need to be error-prone due to the necessity of evolution \cite{Protein_model}. Based on this model, the next step is to consider the mutation process as a transition probability matrix to determine possible mutation directions. Due to the gene sequence's length, obtaining all possible mutations for each nucleotide or amino acid is impractical. Therefore, identifying or extracting mutation-prone positions is crucial.

\vspace{-0.2cm}
\subsection{Identification of Mutation Hot-spots}

\begin{figure}[t]
\begin{center}
  \includegraphics[width=2.57in]{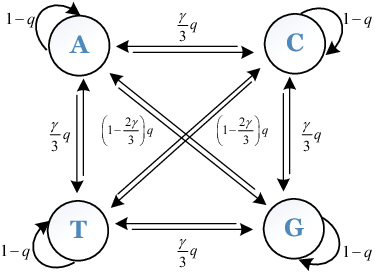}
  \end{center}
   \vspace{-0.25cm}
  \caption{Mutation channel of base substitutions and their mutation probability.}
     \vspace{-0.25cm}
  \label{figure-Mutation channel}
\end{figure}

Depending on the research object, the identification of mutation hot-spots can be divided into two levels, nucleotide or amino acid. Since amino acids can be obtained by translating codons composed of nucleotides, the intrinsic nature of mutation hot-spots is the same, but the scope of the analysis differs.
First, whichever research subjects are considered, the core idea is to search for the mutation-prone position based on the Shannon entropy.
We define $H(i)$ as the entropy of the $i$-th position in the sequence to be analyzed. It is calculated as the sum of $-{{P_k}} (i) \times {\log _2}{P_k}(i)$ for all $k$ in the set $N$, where $P_k(i)$ is the probability of the $k$-th type of nucleotide (or amino acid) at the $i$-th position in the sequence and $N$ is a list of all possible nucleotides (or amino acids).
Taking the nucleotide as an example, if we need to calculate $P_k(i)$ accurately for the considered virus, it is necessary to collect the genetic sequence (or protein sequence) data as much as possible.
The positional entropy described previously is a measure of the randomness at the given position in the sequence. Generally, a higher value of entropy at a position in the sequence suggests the increased randomness at that site, which is a clue to which sites can easily be mutated in gene sequence or protein \cite{Shannon_entropy_biology}.

\vspace{-0.2cm}
\subsection{Determination of Mutation Direction}

%\begin{figure*}[ht]
%\centering
%  \hspace{-0.2cm} \includegraphics[width=5.3in]{OR3a_prediction_new_new.eps}
%  \caption{Numerical results for mutation prediction in the ORF3a protein of SARS-CoV-2. Top subfigure: Identification of mutation hot-spots. Bottom subfigure: Determination of mutation direction in position 57 of the ORF3a protein with $q\in\{10^{-3},10^{-9}\}$ and $\gamma=0.1$. Please note that the 3168 ORF3a protein sequences used in this simulation are downloaded from the NCBI Virus database (https://www.ncbi.nlm.nih.gov), with a sampling period from December 2019 to January 31, 2020.}
%  \label{figure-Virus evolution}
%\end{figure*}

\begin{figure*}[htb]
    \centering
    \subfigure{
        \includegraphics[width=5.51in]{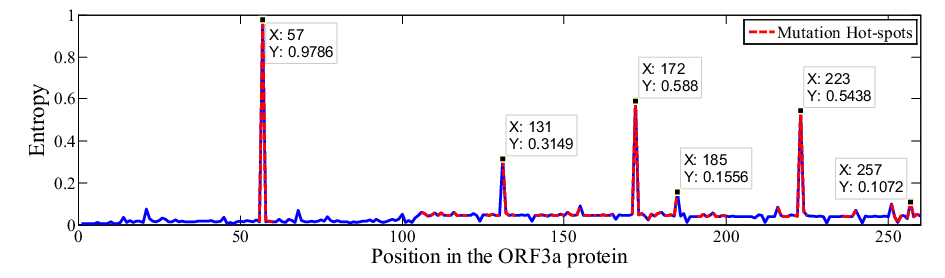}
    }
    \vspace{-0.45cm}
        \subfigure{
	\includegraphics[width=6.13in]{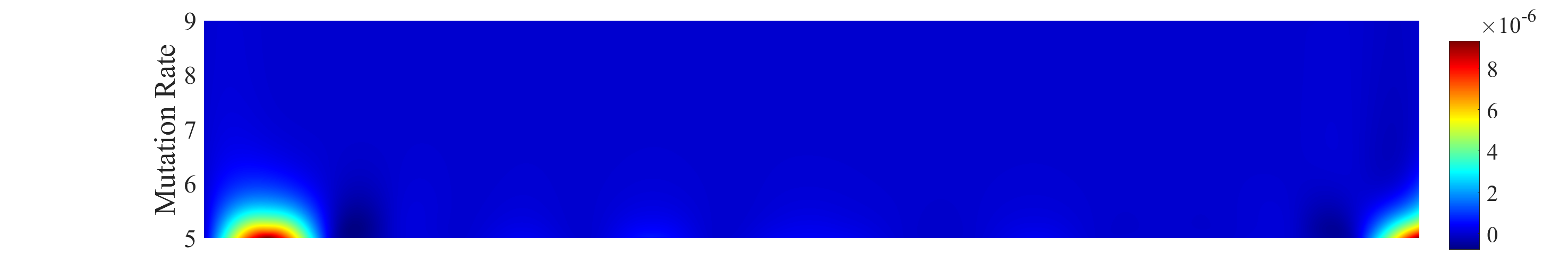}
	}
	    \subfigure{
	\includegraphics[width=6.13in]{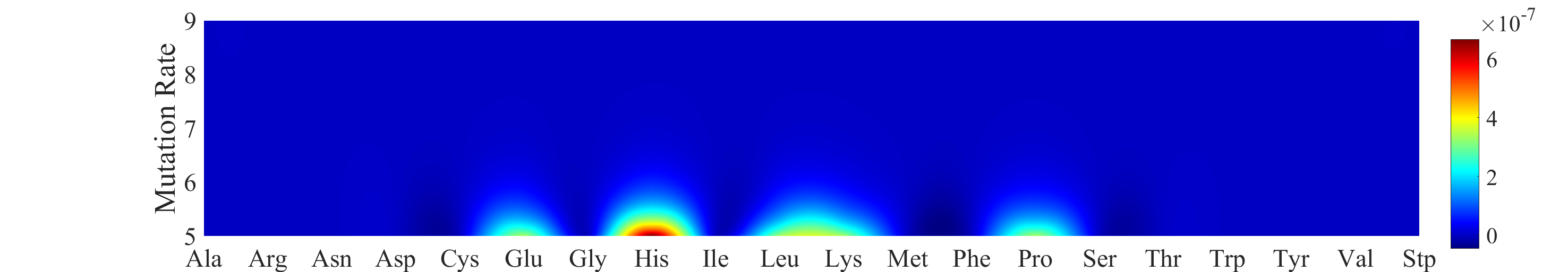}
	}
  \caption{Numerical results for mutation prediction in the ORF3a protein of SARS-CoV-2. Top subfigure: Identification of mutation hot-spots. Bottom subfigure: Determination of mutation direction in position 57 of the ORF3a protein with $q\in\{10^{-3},10^{-9}\}$ and $\gamma=0.1$. Please note that the 3168 ORF3a protein sequences used in this simulation are downloaded from the NCBI Virus database (https://www.ncbi.nlm.nih.gov), with a sampling period from December 2019 to January 31, 2020.}
     \vspace{-0.35cm}
  \label{figure-Virus evolution}
\end{figure*}

The mutation rate can provide valuable insights into the possible evolutionary direction of viruses. For ease of analysis, it is usually assumed that all mutations are mutually independent. Note that in this paper, mutation refers specifically to substitution mutation, i.e., the $k$-th type of nucleotide (or amino acid) is changed as the $k'$-th type of nucleotide (or amino acid) with $k \ne k'$ during the signal transmission. This is because besides substitution, the most relevant mutation with communication theory is insertions and deletions, which cause a synchronization problem at the decoder. Therefore, when considering the mutation of nucleotides, the base substitution channel can be modeled well through the two-parameter Kimura model of nucleotide substitution, which has been extensively used in molecular evolution studies \cite{TIT_DNA_capacity}. For clarity, we illustrate the potential mutation channels for base substitutions and their respective mutation probabilities in Fig.~\ref{figure-Mutation channel}. Here, A, C, T, and G represent the possible nucleotides; $q$ denotes the mutation rate, and $\gamma$ is a factor quantifying the mutation probability between base transitions and base transversions. Meanwhile, a $4 \times 4$ transition probability matrix ${\mathbf{\Pi }}$ for the base substitution channel can be well expressed via Fig.~\ref{figure-Mutation channel} to show the mutation process more concisely. Particularly, the mutation about $\left\{ A \right\} \leftrightarrow \left\{ G \right\}$ and $\left\{ C \right\} \leftrightarrow \left\{ T \right\}$ is base transitions, whereas $\left\{ {A,G} \right\} \leftrightarrow \left\{ {C,T} \right\}$ is the base transversions, due to the chemical structure similarity among compounds in the $\left\{ {A,G} \right\}$ (or $\left\{ {C, T} \right\}$), where $\leftrightarrow $ refers to the mutual substitution between two nucleotide. When a codon (consisting of three nucleotides) is considered, a $64 \times 64$ transition probability matrix ${\mathbf{\Pi }}_c$ can be obtained via the outer products of ${\mathbf{\Pi }}$.
Similarly, if the mutation object is located on the protein level, a $20 \times 20$ transition probability matrix can be calculated. At this point, the blur evolution or mutation direction of the virus can be determined after integrating mutation-prone positions and mutation probabilities based on actual virus sequences. This can provide valuable information for therapeutics and vaccine design as well as insights into blocking the further spread of viruses~\cite{mullick2021understanding}.

To illustrate the potential mutation sites and directions, we plot the Shannon entropy across the ORF3a protein sequence of SARS-CoV-2 and also plot the mutation direction at position 57 of the ORF3a protein in Fig.~\ref{figure-Virus evolution}. Notably, the mutation hot-spots include around 57, 172, and 223, collaborating with results described in \cite{bianchi2021sars}.
Besides, we take position 57 of the ORF3a protein as an example to show the mutation direction. Generally, Glutamine (Q) commonly mutates to Histidine (H) (Q57H). Due to base transition, Glutamine (encoded by CAA and CAG) is likely to mutate to Arginine or a termination codon. While only considering base transversion, one can easily discover from Fig.~\ref{figure-Virus evolution} that Glutamine is most likely to mutate to Histidine, consistent with Q57H.

\vspace{-0.1cm}
\section{Open Challenges and Future Directions}

There are still some open challenges to the application of MC in infectious virus research, which are discussed below. This section also elaborates on some future research directions.

\vspace{-0.1cm}
\subsection{Closed-form Solution to Generalized Channel Model}

%Channel models are important for analyzing viral propagation.
%However, the complex transmission environment makes closed-from channel expression hard to obtain. For example, studying viral spread at the macroscale needs to consider the following factors: temperature, population density, airflow dynamics across enclosed and open areas, etc. For microscale viral spread, the effects of the immune system, viral replication rates, and receptor saturation cannot be ignored. These environmental factors make deriving closed-form channel expressions challenging. Fortunately, an important fact to point out: most viral transmission is based on diffusion. This enables a parametric model designed from Fick's law, where key but uncertain parameters can be obtained via machine learning.

Channel models are crucial for studying viral propagation, yet the complex transmission environment makes obtaining closed-form expressions challenging. For example, studying viral spread at the macroscale needs to consider the following factors: temperature, population density, airflow dynamics across enclosed and open areas, etc. For microscale viral spread, the effects of the immune system, viral replication rates, and receptor saturation cannot be ignored. Despite these challenges, most viral transmission is diffusion-based, allowing for a parametric model derived from Fick's law. Herein, the emerging techniques, such as machine learning or large language models, have the potential to obtain key but uncertain parameters for this model.

\vspace{-0.1cm}
\subsection{Construction of a Microscale Experimental Platform}

Most experiments typically commence at the macroscale rather than the microscale due to the complexity of the nanoscale channel. However, studying virus spread requires mastering communication between infected cells releasing the virus and healthy cells. An experimental model that integrates macro- and micro-scale factors can offer a realistic estimation of infection probability and epidemic timelines. Future endeavors should prioritize the construction of a microscale testbed to simulate virus spread dynamics.

\subsection{Reliable Detection under Complex Communication Blocks}

The uncertainty in the basic modules of MC systems affects detection results, primarily due to the highly dynamic transmitter and the time-varying channel. The varying behavior of infected individuals, such as movement or breathing, leads to an unstable release of the virus. Besides, other biological entities in the same space can greatly impact viral transmission, potentially causing inter-symbol interference. Although threshold-based detection is greatly suitable for OOK with a fixed number of released particles, the actual viral load varies over time and across individuals. Therefore, an effective detection strategy should be designed considering these time variations.

\section{Conclusion}

% Future communication systems, such as 6G, are envisioned to be human-centric, enabling healthcare applications beyond our imagination, such as the transmission/detection modeling and mutation prediction of viruses.

With nanoscale connectivity and high biocompatibility, IoBNT is envisioned to empower epidemic control despite substantial challenges. This paper illustrated how MC presents a solution for the challenges faced in constructing IoBNT for epidemic prevention.
First, we employed MC to provide new insights into the viral propagation process, from which channel models can be obtained. Besides, this method can be used to develop real-time viral detection methods in both macroscale and microscale scenarios, providing a quicker response to emerging pathogens compared to conventional methods. Finally, with the MC system model, virus mutation was analyzed from an information-theoretical point of view, giving a new insight into the prediction of the virus evolution.

\ifCLASSOPTIONcaptionsoff
  \newpage
\fi

\bibliographystyle{IEEEtran}
\bibliography{IEEEabrv,MCvirus}
\vspace{-0.5cm}

\begin{IEEEbiographynophoto}
{Xuan Chen} received her Ph.D. degree from South China University of Technology in 2022. She is currently a Lecturer with Guangzhou University. Her main research interests include molecular communications and index modulation.
\end{IEEEbiographynophoto}
\vspace{-0.25cm}

\begin{IEEEbiographynophoto}
{Yu Huang} received his Ph.D. degree from South China University of Technology in 2021. He is currently an Associate Professor with Guangzhou University. His main research interests include molecular communications and wireless communications.
\end{IEEEbiographynophoto}
\vspace{-0.25cm}

\begin{IEEEbiographynophoto}
{Miaowen Wen} received the Ph.D. degree from Peking University, China, in 2014. He is currently a full-time Professor with South China University of Technology, China. He has published two books and more than 200 IEEE journal articles. His research interests include a variety of topics in the areas of wireless and molecular communications.
\end{IEEEbiographynophoto}
\vspace{-0.25cm}

\begin{IEEEbiographynophoto}
{Shahid Mumtaz} received his M.Sc.
degree from Blekinge Institute of Technology, Sweden, and his
Ph.D. from the University of Aveiro, Portugal. He is currently a Professor at Nottingham Trent University. His research interests include MIMO
techniques, multihop relaying communication, cooperative
techniques, cognitive radios, game theory, and energy-efficient
frameworks for 5G. He has authored several conference papers,
journals, and books.
\end{IEEEbiographynophoto}
\vspace{-0.25cm}

\begin{IEEEbiographynophoto}{Fatih Gulec} received his Ph.D. degree from Izmir Institute of Technology in 2021. He is currently a YUFE MSCA Postdoctoral Fellow with the University of Essex, UK. He held postdoc positions at TU Berlin and York University. His research interests are molecular communications and computational biology. He received the 2022 Doctoral Thesis Award from the IEEE Turkey Section.
\end{IEEEbiographynophoto}
\vspace{-0.25cm}

\begin{IEEEbiographynophoto}
{Anwer Al-Dulaimi} is currently an Associate Professor at Zayed University, UAE, and the Senior Strategy Manager of Connectivity and Industry 4.0 at Veltris, Toronto, Canada. He received his Ph.D. in Electrical and Computer Engineering from Brunel University, London, U.K., in 2012, after obtaining M.Sc. and B.Sc. honors degrees in Communication Engineering. His research interests include 6G networks, cloud computing, V2X, and cybersecurity.
\end{IEEEbiographynophoto}
\vspace{-0.25cm}

\begin{IEEEbiographynophoto}
{Andrew Eckford} received the Ph.D. degree from the University of Toronto, in 2004. He is currently an Associate Professor with York University, Canada. He is a coauthor of the textbook Molecular Communication (Cambridge University Press). His research interest includes the application of information theory to non-conventional channels and systems, especially the use of molecular and biological means to communicate. He was a Finalist for the 2014 Bell Labs Prize.
\end{IEEEbiographynophoto}

\end{document}